\documentclass[12pt]{article}  % list options between brackets
\usepackage{times,amsmath,amsfonts,amsthm,amssymb,eucal}
\usepackage{array}
\usepackage{mathrsfs}
\usepackage{amsmath}
\usepackage{setspace}
\usepackage{fullpage}
\usepackage{mathtools}
\usepackage{fancyhdr}
\usepackage{setspace}
\usepackage{amssymb}
\usepackage{epsfig}
\usepackage{mathrsfs}
\usepackage{mathtools}
\usepackage{enumerate}
\usepackage{float}
\usepackage{graphicx}
\usepackage{makeidx}
\usepackage{color}
\usepackage{multirow}
\usepackage{epsfig}
\usepackage{graphicx}
\usepackage{makeidx}
\usepackage{gensymb}
\usepackage{enumerate}
\usepackage{multirow}
 \usepackage{setspace}
\usepackage{amssymb}
 \usepackage{setspace}
\newtheorem{theorem}{Theorem}
\newtheorem{proposition}{Proposition}
\newtheorem{lemma}{Lemma}
\newtheorem{corollary}{Corollary}
\newtheorem{definition}{Definition}
\def\Theorem{\begin{theorem}\sl}
\def\EndTheorem{\end{theorem}}
\def\Proposition{\begin{proposition}\sl}
\def\EndProposition{\end{proposition}}
\def\Lemma{\begin{lemma}\sl}
\def\EndLemma{\end{lemma}}
\def\Corollary{\begin{corollary}\sl}
\def\EndCorollary{\end{corollary}}
\def\Definition{\begin{definition}\sl}
\def\EndDefinition{\end{definition}}
\numberwithin{equation}{section}

\begin{document}
%%%%%%%%%%%%%%%%%%%%%%%%%%%%%%%%%%%%%%%%%%%%%%%%%%%%%%%%%%%%%%%%%%%%%%%%%%%%%%%%%%%%%%%%%%%%%%%%%%%%%%%%%%%%%%%%%%%%%%%%%%%%
\title{ \textbf{On Simulations form the Two-Parameter Poisson-Dirichlet Process and the Normalized Inverse-Gaussian Process}}   % type title between braces
\author{Luai Al Labadi and Mahmoud Zarepour\thanks{{\em Address for correspondence}: M. Zarepour, Department of Mathematics and Statistics,
University of Ottawa, Ottawa, Ontario, K1N 6N5, Canada. E-mail: zarepour@uottawa.ca.} }        % type author(s) between braces        % type author(s) between braces
\date{\today}    % type date between braces
\maketitle
%%%%%%%%%%%%%%%%%%%%%%%%%%%%%%%%%%%%%%%%%%%%%%%%%%%%%%%%%%%%%%%%%%%%%%%%%%%%%%%%%%%%%%%%%%%%%%%%%%%%%%%%%%%%%%%%%%%%%%%%%%%%
\pagestyle {myheadings} \markboth {} {Al Labadi and Zarepour: Interplay of Frequentist and Bayesian}
\begin{abstract}
In this paper, we develop simple, yet efficient, procedures for sampling approximations of the two-Parameter Poisson-Dirichlet Process and the normalized inverse-Gaussian process. We compare the efficiency of the new approximations to the corresponding stick-breaking approximations of the two-parameter Poisson-Dirichlet Process and the normalized inverse-Gaussian process, in which we demonstrate a substantial improvement.\par

\vspace{9pt} \noindent\textsc{Key words}:  Dirichlet process, Nonparametric Bayesian inference,  Normalized inverse-Gaussian process, Simulation, Stable law process, Stick-breaking representation, Two-parameter Poisson-Dirichlet process.

\vspace{9pt}

\noindent { \textbf{MSC 2000:}} Primary 65C60; secondary 62F15.

\end{abstract}

%%%%%%%%%%%%%%%%%%%%%%%%%%%%%%%%%%%%%%%%%%%%%%%%%%%%%%%%%%%%%%%%%%%%%%%%%%%%%%%%%%%%%%%%%%%%%%%%%%%%%%%%%%%%%%%%%%%%%%%%%%%%
%\setcounter{section}{1}
%\setcounter{equation}{0} %-1
%\begin{center}
%{\textbf{1. INTRODUCTION}}
%\end{center}
\section {Introduction}
\label{intro}

The objective of Bayesian nonparametric inference is to  place a prior on the space of probability measures. The Dirichlet process, formally introduced in Ferguson (1973), is considered the first  celebrated  example on this space. Ferguson's (1973) original definition of the Dirichlet process was based on specifying  its finite-dimensional marginals to be Dirichlet distributions. An alternative definition of the Dirichlet process, due to Ferguson (1973),  was relied on normalizing the gamma process. A different constructive definition of the Dirichlet process was given by Sethuraman (1994) using a ``stick-breaking'' approach. We refer the reader to
the Zarepour and Al Labadi (2012) for more discussion about different representations of the Dirichlet process.

Several alternatives of the Dirichlet process have been proposed in the literature. In this paper, we focus on two such priors, namely the two-parameter Poisson-Dirichlet process (Pitman and Yor, 1997) and the normalized inverse-Gaussian process (Lijoi, Mena and Pr\"unster, 2005).  We begin by introducing the  stick-breaking definition of  the two-parameter Poisson-Dirichlet process defined on an arbitrary measurable space $(\mathfrak{X},\mathcal{A})$. See for more details Pitman and Yor (1997).

\Definition \label{PD1}
For $ 0 \le \alpha <1,$ $\theta>-\alpha,$ let $(\beta_i)_{i \ge 1}$ be a sequence of independent  random
variables with a ${Beta}(1-\alpha,\theta+i\alpha)$ distribution. Define
\begin{equation}
{p'_1}=\beta_1,~ {p'_i}=\beta_i\prod^{i-1}_{k=1}(1-\beta_k),~i\ge2.\nonumber
\end{equation}
Let $p_1 \ge p_2 \ge \ldots$ be the ranked values of $\left({p'_i}\right)_{i\ge 1}$. Moreover, let  $(Y_{i})_{i \ge {1}}$ be a sequence of independent and identically distributed (i.i.d.) random variables with common distribution $H$, independent of $(\beta_i)_{i \ge 1}$.
Then  the random probability measure
\begin{equation}
P_{H,\alpha,\theta}(\cdot)=\sum_{i=1}^{\infty}p_i \delta_{{Y}_i}(\cdot)\label{eq2.888}
\end{equation}
is called a \emph{two-parameter Poisson-Dirichlet process} on $(\mathfrak{X},\mathcal{A})$ with parameters $\alpha,$ $\theta$ and $H,$  where $\delta_x$ denotes the dirac measure at $x$.
\EndDefinition

It is worth mentioning  that Ishwaran and James (2001) referred to the process  ${P}'_{H,\alpha,\theta}(\cdot)=\sum_{i=1}^{\infty}{p'_i} \delta_{{Y}_i}(\cdot)$   as the Pitman-Yor process, where $({p'_i})_{i \ge 1}$ and $(Y_i)_{i \ge 1}$ are as defined in Definition \ref{PD1}. The two-parameter Poisson-Dirichlet process with parameters $\alpha$, $\theta$ and $H$ is denoted by ${PDP}(H;\alpha,\theta)$, and we write
$P_{H,\alpha,\theta}\sim {PDP}(H;\alpha,\theta).$ In the literature, the probability measure $H$ is called the \emph{base measure} of $P_{H,\alpha,\theta}$, while the parameters  $\alpha$  and $\theta$ are  called the \emph{discount parameter}  and  the \emph{concentration parameter}, respectively (Buntine and Hutter, 2010;  Teh, 2006).   The representation (\ref{eq2.888}) clearly shows that any realization of the two-parameter Poisson-Dirichlet process must be a discrete probability measure. Note that,  the special case $PDP(H;0,\theta)$ represents the  Dirichlet process. The law of the weights $(p_1, p_2, \ldots)$ is called the two-parameter Poisson-Dirichlet  distribution, denoted by $PD(\alpha,\theta)$. The two-parameter Poisson-Dirichlet distribution has many applications  in different fields such as  population genetics, ecology, statistical physics and number theory. See Feng (2010) for more details.  On the other hand,  the two-parameter Poisson-Dirichlet process has been   recently used in applications in Bayesian nonparametric statistics such as  computer science (Teh, 2006), species sampling (Jang, Lee and Lee, 2010; Navarrete, Quintana and M\"uller, 2008) and genomics (Favaro, Lijoi, Mena and Pr\"unster, 2009).

The calculations of the moments for the two-parameter Poisson-Dirichlet process are carried out in Carleton (1999). Let $A$ be a measurable subset of $\mathfrak{X}$. Then
\begin{equation}
{E}(P_{H,\alpha,\theta}(A))=H(A) \ \ \text{ and }\ \ {Var}(P_{H,\alpha,\theta}(A))={H(A)(1-H(A))} \frac{1-\alpha}{1+\theta}. \label{aAs.10}
\end{equation}
It follows from (\ref{aAs.10}) that the base measure $H$ plays the role of the center of the process, while both $\alpha$ and $\theta$ control the  variability of $P_{H,\alpha,\theta}$ around $H$. Observe that, for any fixed set  $A \in \mathcal{A}$ and $\epsilon>0,$ we have
\begin{equation}
\label{chey}\Pr\left\{\left|P_{H,\alpha,\theta}(A)-H(A)\right|>\epsilon\right\} \le {H(A)(1-H(A))} \frac{1-\alpha}{(1+\theta)\epsilon^2}.
\end{equation}
Thus, $P_{H,\alpha,\theta}(A) \overset{p} \to H(A)$ as $\theta \to \infty$ (for a fixed $\alpha$) or as $\alpha \to 1,$ $\alpha<1$ (for fixed $\theta$).
In this paper, ``$\overset{p} \to$" and  ``$\overset{a.s.} \to$"  denote  convergence in probability  and  almost sure convergence, respectively.

Analogous  to the Dirichlet process, Lijoi, Mena and Pr\"unster (2005) defined the normalized inverse-Gaussian process ${P}_{H,\theta}=\left\{{P}_{H,\theta}(A)\right\}_{A \in \mathcal{A}}$ by specifying the distribution of $\left({P}_{H,\theta}(A_1), \ldots\,{P}_{H,\theta}(A_{m})\right)$, for a partition $A_1,\ldots,A_m$ of $\mathfrak{X}$, as the  normalized inverse-Gaussian distribution with parameter $ \left(\theta H(A_1), \ldots\,\theta H(A_m)\right)$, where $m\ge 2$. See equation (3) of Lijoi, Mena and Pr\"unster (2005) for the density of the normalized inverse-Gaussian distribution. The normalized inverse-Gaussian process with parameter $\theta$ and $H$ is denoted by $\text{N-IGP}(H;\theta)$, and we write ${P}_{H,\theta}\sim \text{N-IGP}(H;\theta).$

One of the basic properties of the normalized inverse-Gaussian process is that for any   $A \in \mathcal{A},$
  \begin{equation}
{E}({P}_{H,\theta}(A))=H(A) \ \ \text{ and }\ \ {Var}({P}_{H,\theta}(A))=\frac{H(A)(1-H(A))}{\xi(\theta)}, \label{NIG1}
\end{equation}
where here and throughout this paper $$\xi(\theta)=\frac{1}{\theta^2 e^\theta \Gamma(-2,\theta)}$$
and $\Gamma(-2,\theta)=\int_{\theta}^\infty t^{-3}e^{-t}dt$.

Observe that, for large $\theta$, $\xi(\theta) \approx {\theta}$ (Abramowitz and Stegun, 1972, Formula 6.5.32, page 263), where we use the notation $f(\theta) \approx g(\theta)$ if $\lim_{\theta \to \infty} {f(\theta)}/{g(\theta)}=1.$ Like the two-parameter Poisson-Dirichlet process, it follows from (\ref{NIG1}) that $H$ plays the role of the center
 of the process, while  $\theta$  can be viewed as  the concentration parameter. The larger $\theta$ is, the more likely it is that the realization of ${P}_{H,\theta}$ is close to $H$. Specifically, for any fixed set  $A \in \mathcal{A}$ and $\epsilon>0,$ we have ${P}_{H,\theta}(A) \overset{p} \to H(A)$
 as $\theta \to \infty$ since
\begin{equation}
\Pr\left\{\left|{P}_{H,\theta}(A)-H(A)\right|>\epsilon\right\} \le \frac{H(A)(1-H(A))}{\xi(\theta)\epsilon^2}.\label{cheb0}
\end{equation}

Similar to the Dirichlet process, a series representation of the normalized inverse-Gaussian process can be easily derived  from the Ferguson and Klass representation (1972). Specifically, let $(E_i)_{i \ge 1}$ be a sequence of i.i.d. random variables with an exponential distribution with mean of 1. Define
\begin{equation}
\Gamma_i=E_1+\cdots+E_i. \label{eq2}
\end{equation}
Let $(Y_i)_{i \geq 1}$  be a sequence of i.i.d. random variables with values in $\mathfrak{X}$ and common distribution $H$, independent of $(\Gamma_i)_{i
\ge 1}$. Then  the normalized inverse-Gaussian process with parameter $\theta$ and $H$ can be expressed  as a normalized series representation
\begin{equation}
P_{H,\theta}(\cdot)=\sum_{i=1}^{\infty} {\frac{L^{-1}(\Gamma_i)}{\sum_{i=1}^{\infty}{{L^{-1}(\Gamma_i)}}}\delta_{Y_i}(\cdot)},\label{eq3}
\end{equation}
where
\begin{equation}
L(x)=\frac{\theta}{\sqrt{2\pi}} \int_{x}^{\infty}{e^{-t/2}}{t^{-3/2}}~dt, \text{ for } x>0, \label{eq5}
\end{equation}
and $\delta_X$ denotes the Dirac measure at $X$ (i.e. $\delta_X(B)=1$
if $X \in B$ and $0$ otherwise).  Observe that, working with (\ref{eq3}) is difficult in practice because no closed form for the inverse of the L\'evy measure (\ref{eq5}) exists. Moreover, to determine the random weights in (\ref{eq3}) an infinite sum must be computed.

A radically different constructive definition of the normalized inverse-Gaussian process was recently established  by Favaro, Lijoi and Pr\"unster (2012) using a ``stick-breaking''
approach.  Let $(Z_i)_{i \ge 1}$ be i.i.d. random variables with $Z_i$ is 1/2-stable random variable with scale parameter 1. Define a sequence of dependent random variables $(V_i)_{i \ge 1}$  as follows
\begin{equation}
V_1=\frac {X_1} {X_1+Z_1}~ \text{  such that } X_1\sim \text{GIG}(\theta^2,1,-\frac{1}{2}),\label{pi}
\end{equation}
\begin{equation}
V_i|V_1,\ldots,V_{i-1}=\frac {X_i} {X_i+Z_i}~ \text{  such that } X_i\sim \text{GIG}\left(\frac{\theta^2}{\prod_{j=1}^{i-1}(1-V_j)},1,-\frac{1}{2}\right), \ \ i \ge 2,\nonumber
\end{equation}
where the sequences $(X_i)_{i \ge 1}$ and $(Z_i)_{i \ge 1}$ are independent and GIG denotes the generalized inverse Gaussian distribution (see equation (2) of Favaro, Lijoi and Pr\"unster, 2012). Define
\begin{equation}
p_1=V_1,~ p_j=V_j\prod^{j-1}_{i=1}(1-V_i),~j\ge2.\label{pii-IG}
\end{equation}
Moreover, let  $(Y_i)_{i \ge {1}}$ be a sequence of i.i.d. random variables with common distribution $H$, independent of $(V_i)_{i \ge 1}$.
Define
\begin{equation}
P_{H,\theta}(\cdot)=\sum_{i=1}^{\infty}p_i \delta_{Y_i}(\cdot).\label{eq6-IG}
\end{equation}
Then $P_{H,\theta}$ is a normalized inverse-Gaussian process with parameter $\theta$ and $H$.

This paper is organized as follows. In Section 2, we use the stick-breaking representations (\ref{eq2.888}) and (\ref{eq6-IG}) to sample approximations of the two-parameter Poisson-Dirichlet Process and the normalized inverse-Gaussian process, respectively. We also show in this section that  the stick-breaking
representations are  inefficient  for simulation purposes. In Sections 3 and 4, we develop simple, yet efficient, algorithms to simulate approximations of the two-parameter Poisson-Dirichlet Process and the normalized inverse-Gaussian process, respectively.  An extensive simulation study evaluating the accuracy of the new methods to the stick-breaking approximations is presented in Section 4. The simulation results clearly show that the new approximations are more efficient.

\section{Simulation from Stick-Breaking Representation in Nonparametric Bayesian Inference}
Stick-breaking representations are of special interest in  Bayesian nonparametric inference. For example, they are used in modeling Bayesian hierarchical mixture models (Ishwaran and James, 2001; Kottas and Gelfand,  2001). They are also used to compute the moments and  some theoretical properties of the related priors (Carleton, 1999). In this section, we focus on stick-breaking representations of the two-parameter Poisson-Dirichlet process and the   normalized inverse-Gaussian process from computational point of view.

The  stick-breaking representation of the two-parameter Poisson-Dirichlet process (see Definition \ref{PD1}) can be used to  simulate the approximated two-parameter Poisson-Dirichlet process using a truncation argument. By truncating the higher order terms in the sum (\ref{eq2.888}), we can approximate the  stick breaking representation by
\begin{equation}
P_{n,H,\alpha,\theta}(\cdot)=\sum_{k=1}^{n}p_i \delta_{Y_i}(\cdot), \label{approx.PD}
\end{equation}
where $(\beta_i)_{i \ge 1}$, $(p_i)_{i \ge 1}$, and $(\alpha_i)_{i \ge 1}$ are as given by Definition \ref{PD1} with $\beta_n=1$ (hence $\beta_n$ does not have a beta distribution). The assumption that $\beta_n=1$ is necessary to   make the weights add to 1, almost surely (Ishwaran and James, 2001). A random stopping rule for choosing $n=n(\epsilon)$, where $\epsilon \in (0,1)$, is:
\begin{equation}
n=\inf\left\{i:{p}'_i=(1-\beta_1)\ldots(1-\beta_{i-1})\beta_i<\epsilon \right\}.\label{eq800}
\end{equation}
 The random stoping rule in (\ref{eq800}) is similar to the one in (\ref{eq800}) proposed by Muliere and Tradella (1998)  for the Dirichlet process. The following lemma shows that the weights  $(p_i)_{\i \ge 1}$ in the stick-breaking representation are  not strictly decreasing, almost surely (they are only stochastically decreasing). This makes the truncated stick-breaking representation inefficient for simulation purposes.

\Lemma \label{proba2} Let $({p'}_{i})_{i \ge 1}$ be as in Definition \ref{PD1}. Then $\Pr\left\{{p'}_{i+1}<{p'}_{i}\right\}=\int_{0}^1 \int_{0}^yf(x,y) dx dy$,
where
\begin{equation}
f(x,y)=\frac{x^{\alpha_1-1}(1+x)^{-\alpha_1-\beta_1}}{B(\alpha_1,\beta_1)}\times \frac{y^{\alpha_1-1}(1-y)^{\beta_2-1}}{B(\alpha_1,\beta_2)}I\{x \ge 0\}I\{0<y<1\},\nonumber
\end{equation}
$B(a,b)=\Gamma(a)\Gamma(b)/\Gamma(a+b)$, $\alpha_1=1-\alpha$, $\beta_1=\theta+i\alpha$ and $\beta_2=\theta+(1+i)\alpha$.
\EndLemma

\proof
Since ${p'_i}=\beta_i\prod^{i-1}_{k=1}(1-\beta_k),$ we have
\begin{eqnarray}
\nonumber\Pr\left\{{p'}_{i+1}<{p'}_{i}\right\}&=&\Pr\left\{\beta_{i+1}(1-\beta_{i})<\beta_{i}\right\}\\
\nonumber &=&\Pr\left\{\beta_{i+1}\frac{(1-\beta_{i})}{\beta_i}<1\right\}.
\end{eqnarray}
Since $\beta_i$ is a random variable with the ${Beta}(1-\alpha,\theta+i\alpha)$ distribution, it follows that $\beta_i/(1-\beta_i)$ has the beta distribution of the second kind  with parameters $\alpha_1=1-\alpha$ and $\beta_1=\theta+i\alpha$ (Balakrishnan and Lai, 2009, page 12). That is, $\beta_i/(1-\beta_i)$ has the density
$$f(x)=\frac{x^{\alpha_1-1}(1+x)^{-\alpha_1-\beta_1}}{B(\alpha_1,\beta_1)}I\{x \ge 0\}.$$ The lemma follows from the fact that $(\beta_i)_{i \ge 1}$ is a sequence of independent  random variables with a $\text{Beta}(1-\alpha,\theta+i\alpha)$ distribution.
\endproof
It follows clearly from Lemma \ref{proba2} that the probability $\Pr\left\{{p}'_{i+1}<{p}'_{i}\right\}$ depends on $i,$ $\alpha$ and $\theta$. Table 1 depicts some values for this probability.

\begin{table}[h!]
\caption{Some values of $\Pr\left\{{p}'_{i+1}<{p}'_{i}\right\}$.}
\begin{center}
\begin{tabular}{|l|c|c|c|}
\hline
$i$& $\alpha$ & $\theta$ &  $\Pr\left\{{p}'_{i+1}<{p}'_{i}\right\}$ \\
\hline
1& 0.1&1& 0.672\\
\hline
10& 0.1&1& 0.607\\
\hline
100& 0.1&1&0.521 \\
\hline
1& 0.5&1& 0.598\\
\hline
10& 0.5&1& 0.526\\
\hline
100& 0.5&1& 0.503\\
\hline
1& 0.9&1& 0.5230\\
\hline
10& 0.9&1&0.504\\
\hline
100& 0.9&1&0.500 \\
\hline
1& 0.1&10&  0.523\\
\hline
10& 0.1&10& 0.521\\
\hline
100& 0.1&10&0.511 \\
\hline
1& 0.5&10&0.515 \\
\hline
10& 0.5&10& 0.511\\
\hline
100& 0.5&10& 0.503\\
\hline
1& 0.9&10& 0.504\\
\hline
10& 0.9&10& 0.502\\
\hline
100& 0.9&10&0.500     \\
\hline
\end{tabular}
\end{center}
\end{table}

Similar to  the two-parameter Poisson-Dirichlet process, the stick-breaking representation of  the normalized inverse-Gaussian process can be used to approximately simulate the normalized inverse-Gaussian process process using a truncation argument. By truncating the higher order terms in the sum (\ref{eq6-IG}), we can approximate the  stick breaking representation by
\begin{equation}
P_{n,H,\theta}(\cdot)=\sum_{k=1}^{n}p_i \delta_{Y_i}(\cdot), \label{approx.PD-IG}
\end{equation}
where $(V_i)_{i \ge 1}$, $(p_i)_{i \ge 1}$ are as given by Definition \ref{PD1} with $V_n|V_1,\ldots,V_{n-1}=1$. The assumption that $V_n|V_1,\ldots,V_{n-1}=1$ is necessary to   make the weights add to 1, almost surely. A random stopping rule for choosing $n=n(\epsilon)$, where $\epsilon \in (0,1)$, is similar to (\ref{eq800}) with $\beta_i$ is replaced by $V_i$. Note that, since the
$(V_i)_{i\ge 1}$ are not independent and the joint density of $(V_i,V_{i+1})$ is complex for a direct calculation, establishing a lemma similar to Lemma \ref{proba2} for the normalized inverse-Gaussian process is not an easy task. The next table gives values of $\Pr\left\{{p}_{i+1}<{p}_{i}\right\}$ based on simulation where, the values of the probability is based on 500 simulated values  of each weight with $\theta=1$ and $n=50$.

\begin{table}[h!]
\caption{Some values of $\Pr\left\{{p}_{i+1}<{p}_{i}\right\}$.}
\begin{center}
\begin{tabular}{|l|c|c|}
\hline
$i$& $\alpha$ & $\Pr\left\{{p}_{i+1}<{p}_{i}\right\}$ \\
\hline
1& 1&0.536\\
\hline
10& 1& 0.538\\
\hline
20& 1&0.540 \\
\hline
30& 1&0.548\\
\hline
40& 1& 0.664\\
\hline
\end{tabular}
\end{center}
\end{table}

\section{Simulating an Approximation of the Two-Parameter Poisson-Dirichlet Process}

The next proposition provides an interesting approach to construct the two-parameter Poisson-Dirichlet process. For the proof of the proposition, see Pitman and Yor (1997, Proposition 22).
\Proposition \label{PD-sim}
For $0<\alpha<1$ and $\theta>0$, suppose  $\left(p_1(0,\theta),p_2(0,\theta),\ldots \right)$ and $(p_1(\alpha,0),$ $p_2(\alpha,0),\ldots)$  has respective distributions $PD(0,\theta)$ and $PD(\alpha,0)$. Independent of $(p_1(0,\theta)$, $p_2(0,\theta),\ldots )$, let $\left(p^i_1(\alpha,0),p^i_2(\alpha,0),\ldots\right), i=1,2,\ldots,$ be  a sequence of independent copies of $\left(p_1(\alpha,0),p_2(\alpha,0),\ldots\right)$. Let $(p_i)_{i\ge 1}$ be the descending order statistics of $\{p_i(0,\theta)p_j^{i}(\alpha,0),$ $i,j=1,2,\ldots\}$. Then $\left(p_1,p_2,\ldots\right)$ has a $PD(\alpha,\theta)$ distribution.
\EndProposition

It follows form  Proposition \ref{PD-sim}, that the weights in the two-parameter Poisson-Dirichlet process can be
constructed based on  two boundary selections of the parameters. The first selection is when $\alpha=0$. This choice of parameters
corresponds to the Dirichlet process. The other selection of parameters is when $\theta=0$, which yields a measure whose random weights are based on a stable law with index $0<\alpha<1$. Therefore, the Dirichlet process $P_{H,0,\theta}$ and the  stable law process $P_{H,\alpha, 0}$ are two essential processes in simulating the two-parameter Poisson-Dirichlet process. First we consider simulating these two key processes.

A simple, yet efficient, procedure for approximating the Dirichlet process was was recently developed  by Zarepour and Al Labadi (2012). Specifically, let $X_n$  be a random variable with distribution $\text{Gamma}(\theta/n,1)$. Define
\begin{equation}
G_n(x)=\Pr(X_n>x)=\int_{x}^{\infty}{\frac{1}{\Gamma(\theta/n)}e^{-t}t^{\theta/n-1}dt}. \label{eq9}
\end{equation}
and
\begin{equation}
G^{-1}_n(y)=\inf\left\{x:G_n(x)\ge y\right\}. \nonumber
\end{equation}
Let $(Y_i)_{i \geq 1}$ be a sequence of i.i.d. random variables
with values in $\mathfrak{X}$ and common distribution $H$, independent
of $(\Gamma_i)_{i \ge 1}$. Let  $\Gamma_i=E_1+\cdots+E_i,$ where $(E_i)_{i\ge 1}$ are i.i.d. random variables with exponential distribution of
 mean 1, independent of $\left(Y_i\right)_{i \ge 1}$
Then as $n \to \infty$,
\begin{equation}
P^{\text{new}}_{n,H,0,\theta}(\cdot)=\sum_{i=1}^{n} \frac
{{G_n^{-1}\left(\frac{\Gamma_i}{\Gamma_{n+1}}\right)}}{\sum_{i=1}^{n}{G_n^{-1}\left(\frac{\Gamma_i}{\Gamma_{n+1}}\right)}}\delta_{Y_i}(\cdot)\overset{a.s.}\rightarrow \sum_{i=1}^{\infty} {\frac{N^{-1}(\Gamma_i)}{\sum_{i=1}^{\infty}{{N^{-1}(\Gamma_i)}}}\delta_{Y_i}}(\cdot),
\label{eq11}
\end{equation}
where $N(x)=\theta\int_{x}^{\infty}t^{-1}e^{-t}dt,~x>0$. The right-hand side of (\ref{eq11}) represents  Ferguson's representation (1973) of the Dirichlet process.  Zarepour and Al Labadi (2012) showed that the weights ${{G_n^{-1}\left(\frac{\Gamma_i}{\Gamma_{n+1}}\right)}}/{\sum_{i=1}^{n}{G_n^{-1}\left(\frac{\Gamma_i}{\Gamma_{n+1}}\right)}}$ of the new representation given in (\ref{eq11}) decrease monotonically for any fixed positive integer $n$. They also  provided a strong empirical evidence that their new representation yields a highly accurate approximation of the Dirichlet process.

The next algorithm uses Zarepour and Al Labadi approximation (\ref{eq11}) to generate a sample from the approximate Dirichlet process with parameters $\theta$ and $H$.

\vspace{3mm}
\noindent \textbf{Algorithm A: Simulating an approximation of the  Dirichlet process.}
\begin{enumerate} [(1)]
\item Fix a relatively large positive integer $n$
.
\item Generate  $Y_i\overset{\text{i.i.d.}}\sim H$ for $i=1,\ldots,n.$
\item For $i=1,\ldots,n+1,$ generate $E_i$ from an exponential distribution with  mean 1, independent of $\left(Y_i\right)_{1\le i \le n}$ and let $\Gamma_i=E_1+\cdots+E_i.$
\item For $i=1,\ldots,n,$ compute $G_n^{-1}\left({\Gamma_i}/{\Gamma_{n+1}}\right),$  which is  simply the quantile function of the ${Gamma}(\theta/n,1)$ distribution evaluated at  $1-{\Gamma_i}/{\Gamma_{n+1}}.$
\item Set $P^{\text{new}}_{n,H,0,\theta}$ as defined in (\ref{eq11}) .
 \end{enumerate}

On the other hand, for the stable law process, Pitman and Yor (1997, Proposition 10) proved that
\begin{equation}
P_{H,\alpha,0}(\cdot)=\sum_{i=1}^\infty \frac{\Gamma_i^{-1/\alpha}}{\sum_{i=1}^{\infty}\Gamma_i^{-1/\alpha}}\delta_{Y_i}(\cdot), \label{stable}
\end{equation}
where $\Gamma_i=E_1+\cdots+E_i$ and $(E_i)_{i \ge 1}$ is  a sequence of i.i.d. random variables with an exponential distribution with mean of 1.
Therefore, the following representation can be used to simulate an approximation of the  stable law process
\begin{equation}
P_{n,H,\alpha,0}(\cdot)=\sum_{i=1}^n \frac{\Gamma_i^{-1/\alpha}}{\sum_{i=1}^{n}\Gamma_i^{-1/\alpha}}\delta_{Y_i}(\cdot), \label{stable1}
\end{equation}

It is easy to see that the weights $\left({\Gamma_i^{-1/\alpha}}/{\sum_{i=1}^{n}\Gamma_i^{-1/\alpha}}\right)_{1\le i\le n}$ are strictly decreasing. Thus, simulating the  stable law process through the representation (\ref{stable1}) is very efficient. The next algorithm can be used to  sample from an approximation of the  stable law process.

 \vspace{3mm}

\noindent \textbf{Algorithm B: Simulating an approximation of the  stable law process.}
\begin{enumerate} [(1)]
\item Fix a relatively large positive integer $n$.
\item Generate  $Y_i\overset{\text{i.i.d.}}\sim H$ for $i=1,\ldots,n.$
\item For $i=1,\ldots,n+1,$ generate $E_i$ from an exponential distribution with  mean 1, independent of $\left(Y_i\right)_{1\le i \le n}$ and let $\Gamma_i=E_1+\cdots+E_i.$
\item For each $i=1,\ldots,n,$ the corresponding weights are $\Gamma_i^{-1/\alpha}/\sum_{i=1}^{n}\Gamma_i^{-1/\alpha}$.
 \end{enumerate}

Now we present an efficient algorithm for simulating the two-parameter Poisson-Dirichlet process. This algorithm is based on Proposition \ref{PD-sim}, Algorithm A and Algorithm B.

\vspace{3mm}

\noindent \textbf{Algorithm C: Simulating an approximation of the two-parameter Poisson-Dirichlet Process.}
\begin{enumerate} [(1)]
\item Use Algorithm A to generate $n$ weights of the Dirichlet process. Denote these weights by $\left(p_1(0,\theta),\ldots,p_2(0,\theta)\right)$ .
\item Use Algorithm B to generate $m$ weights for an approximation of the  stable law process. Denote these weights by  $\left(p_1(\alpha,0),\ldots,p_m(\alpha,0)\right)$.
\item Repeat step (2) to generate $n$ i.i.d. copies of $\left(p_1(\alpha,0),\ldots,p_m(\alpha,0)\right)$. Denotes these copies by  $\left(p^1_1(\alpha,0),\ldots,p^1_m(\alpha,0)\right),\ldots,\left(p^n_1(\alpha,0),\ldots,p^n_m(\alpha,0)\right)$.
\item Find the product $p_i(0,\theta)p_j^i(\alpha,0)$ of the weights generated in step (1) and step (3), where $i=1, \ldots, n$ and  $j=1, \ldots, m$. That is, find
$(p_1(0,\theta)p^1_1(\alpha,0),\ldots,$ $p_1(0,\theta)p^1_m(\alpha,0),\ldots,$ $p_n(0,\theta)p^n_1(\alpha,0),\ldots,p_n(0,\theta)p^n_m(\alpha,0))$.
\item The weights of the two-parameter Poisson-Dirichlet process are those weights obtained in step (4) written in descending order. Denote these weight by $(p_i)_{1 \le i \le nm}$.
\item  Generate  $Y_i\overset{\text{i.i.d.}}\sim H$ for $i=1,\ldots,nm.$
\item The approximated two-parameter Poisson-Dirichlet process is given by the representation (\ref{approx.PD}) with $n$ in the summation replaced by $nm$.
 \end{enumerate}

\section{Monotonically Decreasing Approximation to the Normalized Inverse-Gaussian Process}
Mimicking Theorem 1 of  Zarepour and Al Labadi (2012), we can construct a similar approximation for the normalized inverse-Gaussian process. Specifically, let $X_n$  be a random variable with distribution $\text{IG}(\theta/n,1)$. Define
\begin{equation}
Q_n(x)=\Pr(X_n>x)=\int_{x}^{\infty}\frac{\theta}{n\sqrt{2\pi}}t^{-3/2}\exp\left\{-\frac{1}{2}\left(\frac{\theta^2}{n^2t}+t\right)+\frac{\theta}{n}\right\}dt. \label{eq9-IG}
\end{equation}
Let $(Y_i)_{i \geq 1}$ be a sequence of i.i.d. random variables
with values in $\mathfrak{X}$ and common distribution $H$, independent
of $(\Gamma_i)_{i \ge 1}$, then as $n \to \infty$
\begin{equation}
P^{\text{new}}_{n,H,0,\theta}(\cdot)=\sum_{i=1}^{n} \frac
{{Q_n^{-1}\left(\frac{\Gamma_i}{\Gamma_{n+1}}\right)}}{\sum_{i=1}^{n}{Q_n^{-1}\left(\frac{\Gamma_i}{\Gamma_{n+1}}\right)}}\delta_{Y_i}(\cdot)\overset{a.s.}\rightarrow \sum_{i=1}^{\infty} {\frac{L^{-1}(\Gamma_i)}{\sum_{i=1}^{\infty}{{L^{-1}(\Gamma_i)}}}\delta_{Y_i}}(\cdot).
\label{eq11-IG}
\end{equation}
Here $\Gamma_i$, $L(x)$,  and $Q_n(x)$, are defined in (\ref{eq2}),
(\ref{eq5}), and (\ref{eq9-IG}), respectively.

\vspace{3pt}
\noindent {\textbf{Remark 1.}}  For any $1 \le i \le n,$  ${\Gamma_{i}}/{\Gamma_{n+1}}<{\Gamma_{i+1}}/{\Gamma_{n+1}}$  almost surely. Since $Q_n^{-1}$ is a decreasing function, we have $Q_n^{-1}\left({\Gamma_{i}}/{\Gamma_{n+1}}\right)>Q_n^{-1}\left( {\Gamma_{i+1}}/{\Gamma_{n+1}}\right)$ almost surely. That is, the weights of the new representation (\ref{eq11-IG})  decrease monotonically for any fixed positive integer $n$. As demonstrated in Zarepour and Al Labadi (2012) for the Dirichlet process, we also anticipate that this new representation will yield highly  accurate approximations to the normalized inverse-Gaussian process.

\vspace{5pt}

\noindent \textbf{Algorithm D: Simulating an approximation of the normalized inverse-Gaussian process.}

\begin{enumerate} [(1)]
\item Fix a relatively large positive integer $n$.
\item Generate  $Y_i\overset{\text{i.i.d.}}\sim H$ for $i=1,\ldots,n.$
\item For $i=1,\ldots,n+1,$ generate $E_i$ from an exponential distribution with  mean 1, independent of $\left(Y_i\right)_{1\le i \le n}$ and let $\Gamma_i=E_1+\cdots+E_i.$
\item For $i=1,\ldots,n,$ compute $Q_n^{-1}\left({\Gamma_i}/{\Gamma_{n+1}}\right),$  which is  simply the quantile function of the inverse-Gaussian distribution with parameter $a/n$ and $1$ evaluated at  $1-{\Gamma_i}/{\Gamma_{n+1}}.$ Computing such values is straightforward in R. For example, one may use the package ``GeneralizedHyperbolic".
 \end{enumerate}

\section{Empirical Results: A Comparison with the Stick-breaking Approximation}
In this section, we compare the new approximation of the two-parameter Poisson-Dirichlet process (Algorithm C) and the new approximation of the normalized inverse-Gaussian process (Algorithm D) with the corresponding stick-breaking approximations given in (\ref{approx.PD}) and (\ref{approx.PD-IG}). First we consider the two-parameter Poisson-Dirichlet process. In the simulation, we set $n=100$, $m=500$ in Algorithm C and $n=100\times500=50000$ in (\ref{approx.PD}). We take $H$ to be the uniform distribution on $[0,1].$ We generate 1000 sample paths from the two-parameter Poisson-Dirichlet for different values of $\alpha$ and $\theta$ by using the two approximations. The sample means of generated processes at $x=0.1,0.2,\ldots,0.9,1.0$ are compared with the true mean of the two-parameter Poisson-Dirichlet process $H(x)=x$. Table 3 shows the  maximum mean error (the absolute maximum of the differences between the sample means and the  true means).  For instance, for $\alpha=0.9$ and $\theta=10$, the  maximum mean error is $0.00245$ in the new approach, while it is $0.01149$ in the stick-breaking approximation. Similarly, sample standard deviation  and the population standard deviation can be compared and their maximum errors are reported in Table 3. It is clear from the simulation results in Table 3  that both the  maximum mean  error and the maximum standard deviation error in the new approach are  smaller than that obtained by the stick-breaking approximation. Thus, empirically, simulating the two-parameter Poisson-Dirichlet process by using the  the new approximation (Algorithm C) gives very accurate results.

\begin{table}[H]
%\vspace{-10pt}
\caption{This table reports the  maximum mean (max. mean) error and  maximum standard deviation (max. sd.) error. That is, the absolute maximum difference between the approximated mean (standard deviation) and the actual mean (standard deviation) of $P_{H,\alpha,\theta}(x)=P_{H,\alpha,\theta}((-\infty,x])$ evaluated at $x=0.1,0.2,\ldots,0.9,1.0,$ where  $H$ is a uniform distribution on $[0,1].$}
\begin{center}
\begin{tabular}{lllllll}
\hline
\hline
& &\multicolumn{2}{c}{ New }& &\multicolumn{2}{c}{Stick-breaking}\\
\cline{3-4} \cline{6-7}
\multicolumn{0}{c} {$\alpha$} & \multicolumn{0}{c} {$\theta$} &\multicolumn{0}{c} {max. mean error}  &
\multicolumn{0}{c} {max. sd error} & &\multicolumn{0}{c} {max. mean error}  &
\multicolumn{0}{c} {max. sd error}\\
\hline
0.1& 1& 0.01155& 0.67082 &  & 0.02332&0.67082 \\
\\
0.5& 1& 0.00983& 0.50000 && 0.01334 & 0.50000\\
\\
0.9& 1&0.00564 & 0.22361 & & 0.01303 & 0.22361\\
\\
0.1& 10&0.00993 & 0.28604&& 0.01097 & 0.28604\\
\\
0.5& 10& 0.00368& 0.21320 &&0.00455  &0.21320 \\
\\
0.9& 10&0.00245 &  0.09535 &&  0.01149& 0.13863\\
\\
0.1& 50&0.00236 &0.13284  &&0.00341  &0.13284 \\
\\
0.5& 50& 0.00131& 0.09901 &&0.00227  &0.09901 \\
\\
0.9& 50& 0.00094&  0.04428&&  0.00974 & 0.05468\\
\\
\hline
\end{tabular}
\end{center}
\label{table4.3}
\vspace{-10pt}
\end{table}

Figures 1, 2 and 3 show sample paths for the approximate two-parameter Poisson-Dirichlet process with a uniform distribution on [0,1] as a base measure with different concentration and discount parameters. Distinctly, the new approximation performs very well in all cases. On the other hand, a clear disadvantage of the stick-breaking approximation appears  when $\alpha$ is close to 1 $(\alpha<1)$. In this case, as seen in  Figures  2 and 3, contrary to our anticipation (see inequality (\ref{chey})), the two-parameter Poisson-Dirichlet process is not in the proximity of the base measure.
Thus, the stick-breaking representation performs very poorly when $\alpha$ is close to 1 $(\alpha<1)$.

\begin{figure}[h!]
\centering
\includegraphics[width=1.0\textwidth,height=17cm]{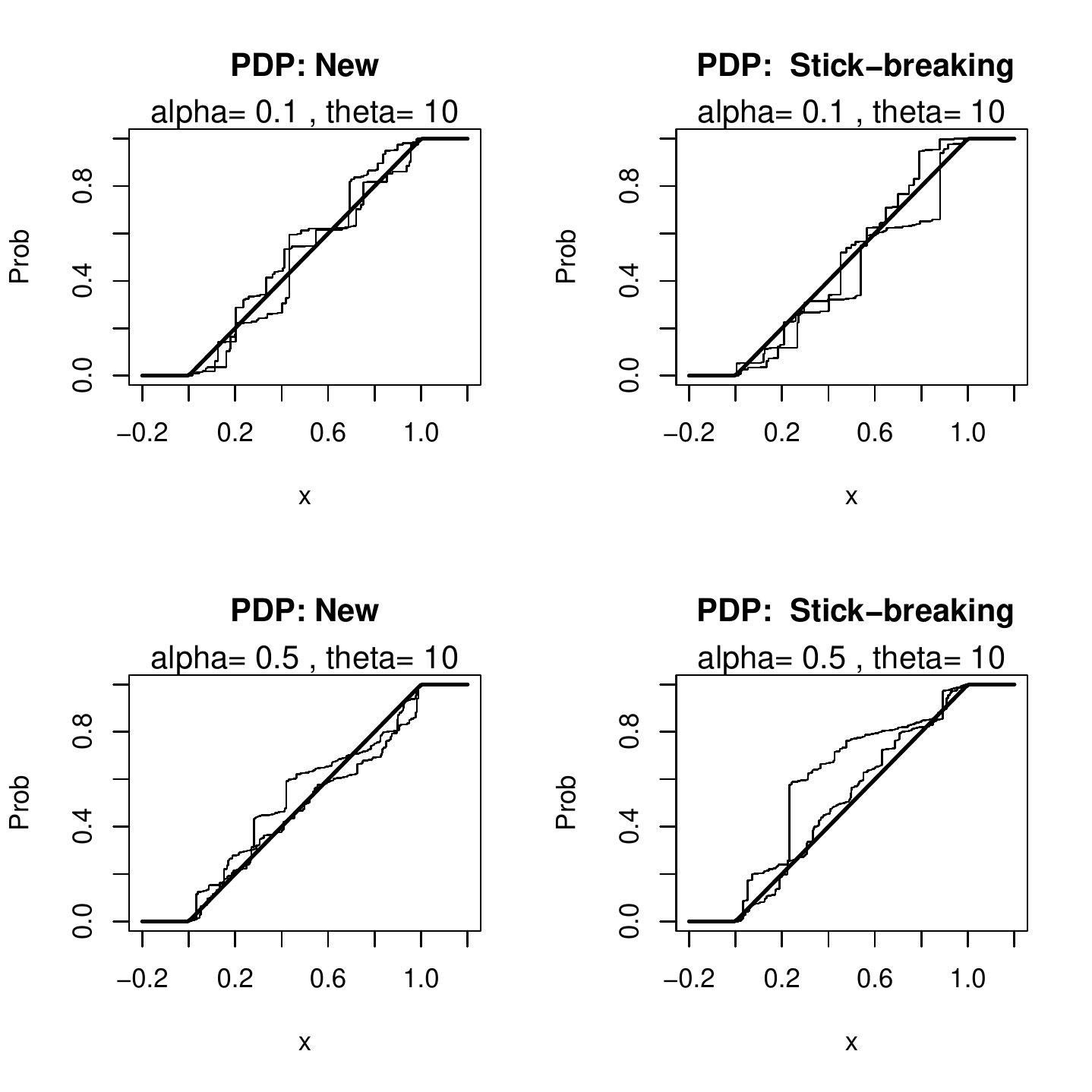}
\caption{Sample paths of the two-parameter Poisson-Dirichlet  process $P_{H,\alpha,\theta}$, where  $H$ is the uniform distribution on $[0,1]$, $\theta=10$ and $\alpha=0.1,0.5$. The solid line denotes the  cumulative distribution function  of $H$.}
\end{figure}

\begin{figure}[h!]
\centering
\includegraphics[width=1.0\textwidth,height=17cm]{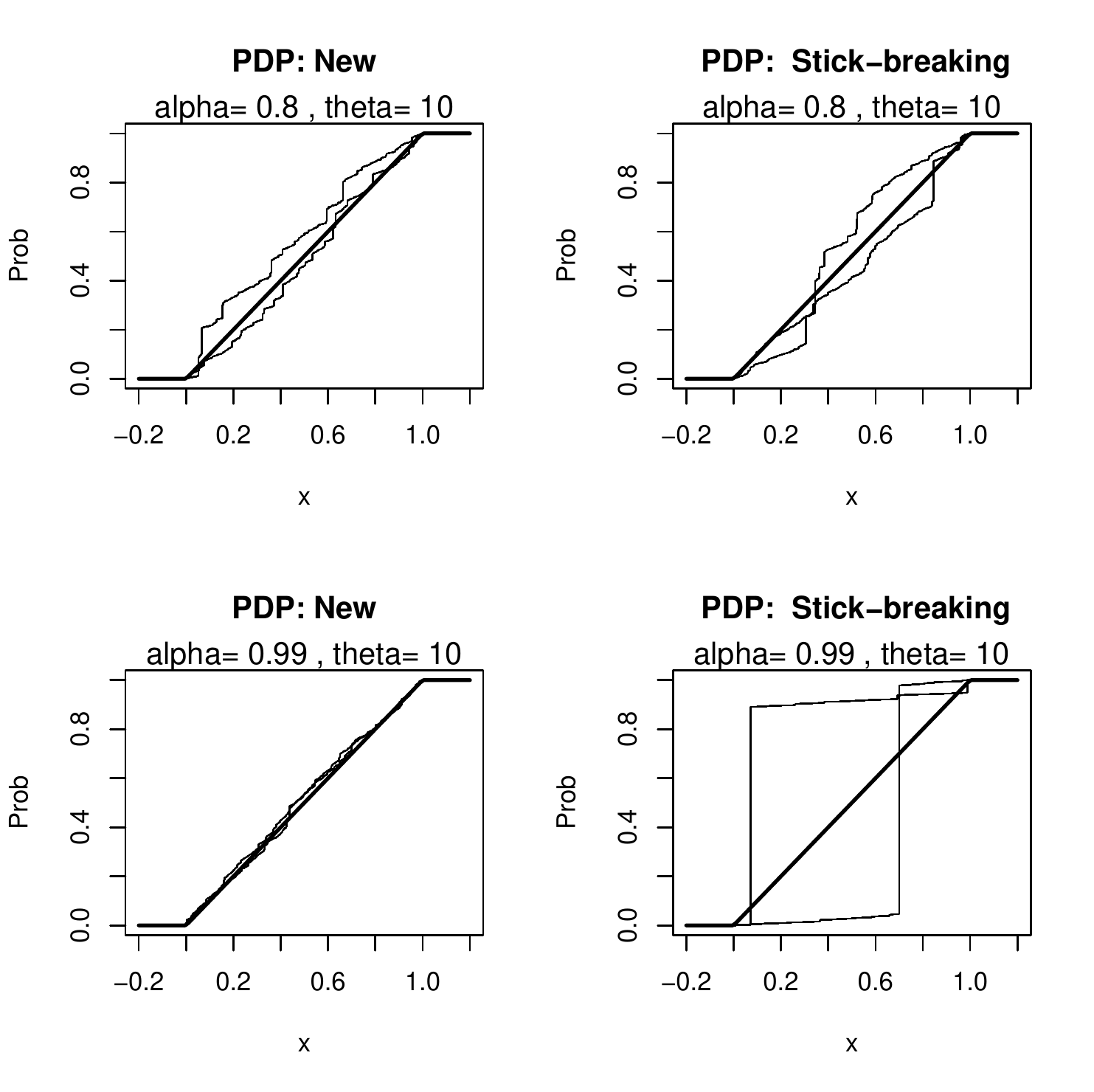}
\caption{Sample paths of a two-parameter Poisson-Dirichlet  process $P_{H,\alpha,\theta}$, where  $H$ is the uniform distribution on $[0,1]$, $\theta=10$ and $\alpha=0.8,0.99$. The solid line  denotes the  cumulative distribution function  of $H$.}
%\vspace{30pt}
\end{figure}

\begin{figure}[h!]
\centering
\includegraphics[width=1.0\textwidth,height=17cm]{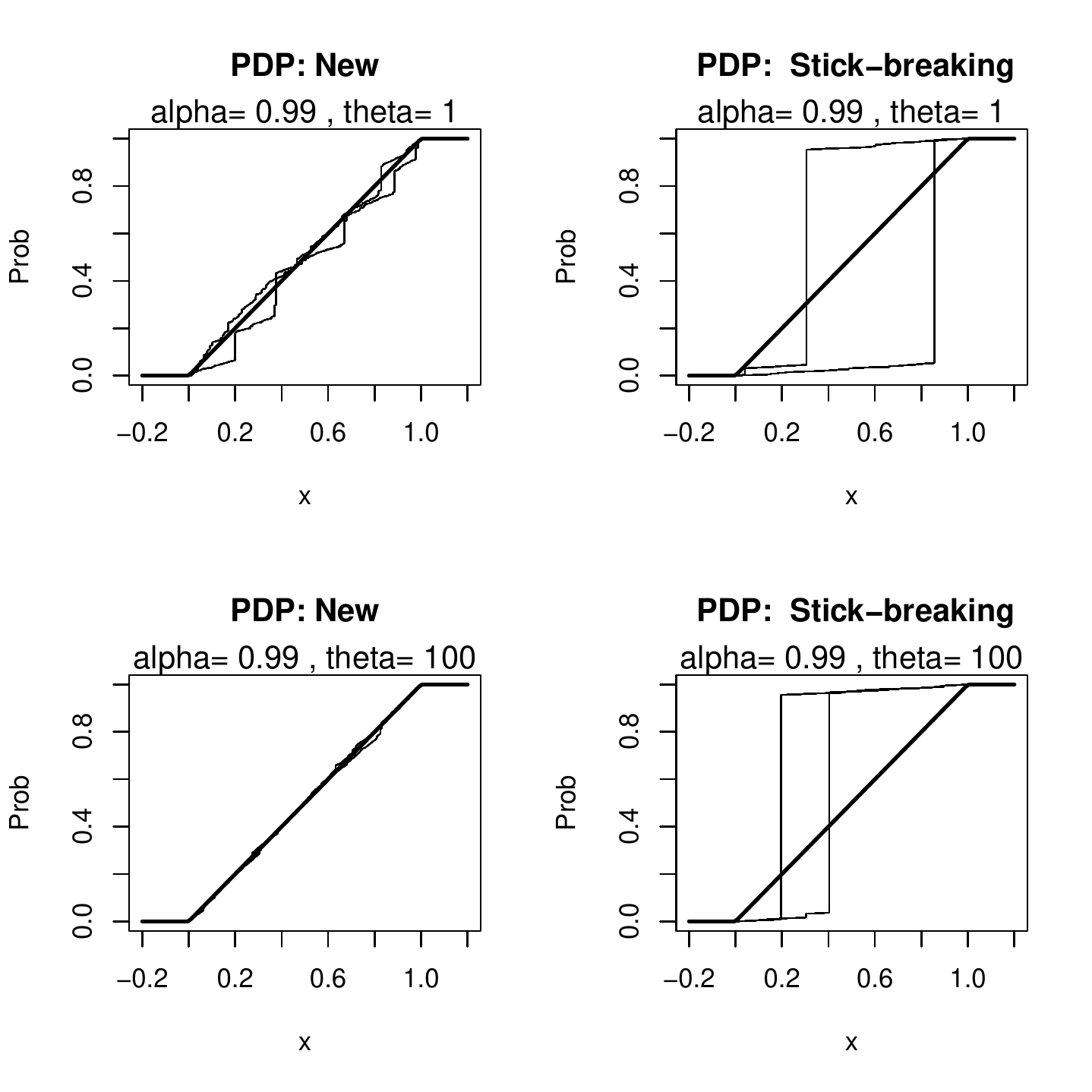}
\caption{Sample paths of a two-parameter Poisson-Dirichlet  process $P_{H,\alpha,\theta}$, where  $H$ is the uniform distribution on $[0,1]$, $\theta=100,1$ and $\alpha=0.99$. The solid line  denotes the  cumulative distribution function  of $H$.}
%\vspace{30pt}
\end{figure}

Next we compare the new approximation of the normalized inverse-Gaussian process (Algorithm D) with the stick-breaking approximation
 (\ref{approx.PD-IG}). In the simulation, we set $n=50$, $\theta=1$ and  take $H$ to be the uniform distribution on $[0,1].$ We generate 500 sample paths from
 the normalized inverse-Gaussian process by using the two approximations. The sample means of generated processes at $x=0.1,0.2,\ldots,0.9,1.0$ are compared with the true mean of the normalized inverse-Gaussian process $H(x)=v$. The  maximum mean error (the absolute maximum of the differences between the sample means and the  true means) is $0.01032$, while it is  $0.01804$ in the stick-breaking approximation. Similarly, the maximum standard deviation
error is $0.06137$ in the new approach, while it is $0.07079$ in the stick-breaking approximation. Once again, both the  maximum mean  error and the  maximum standard deviation error in the new approach are
smaller than those obtained by the stick-breaking approximation.

Figure 4 shows sample paths for the approximate normalized inverse-Gaussian process  with a uniform distribution on [0,1] as a base measure with different concentration parameters. As seen in this figure, the new approximation performs very well in all case.
On the other hand, when $\theta$ is large, contrary to our anticipation (see inequality (\ref{cheb0})), the
normalized inverse-Gaussian process  is not in the proximity of the base measure.

\vspace{3pt}
\noindent {\textbf{Remark 2.}} When generating samples from the approximate stick-breaking representation of
the normalized inverse-Gaussian process, the values of the random variables
 $(V_i)_{i\ge1}$  in (\ref{pi}) become numerically 1 after  few simulation steps ($n \le 50$). The reason of that,
 the values of the random variables $(X_i)_{i\ge1}$ becomes very huge comparable to
 values of the random variables $(Z_i)_{i\ge 1}$. In this case, generating samples for the random variable $(X_i)_{i \ge 1}$
 from the generalized inverse distributions is impossible as these distributions become  undefined  (one of the parameters becomes numerically 0). This problem makes sampling the normalized inverse-Gaussian process via the stick-breaking representation fails in most cases. Our recommendation is to use
 the stick breaking representation to simulate the normalized inverse-Gaussian process only when $\theta$ is very small ($\theta \le 1$).

\begin{figure}[h!]
\centering
\includegraphics[width=1.0\textwidth,height=17cm]{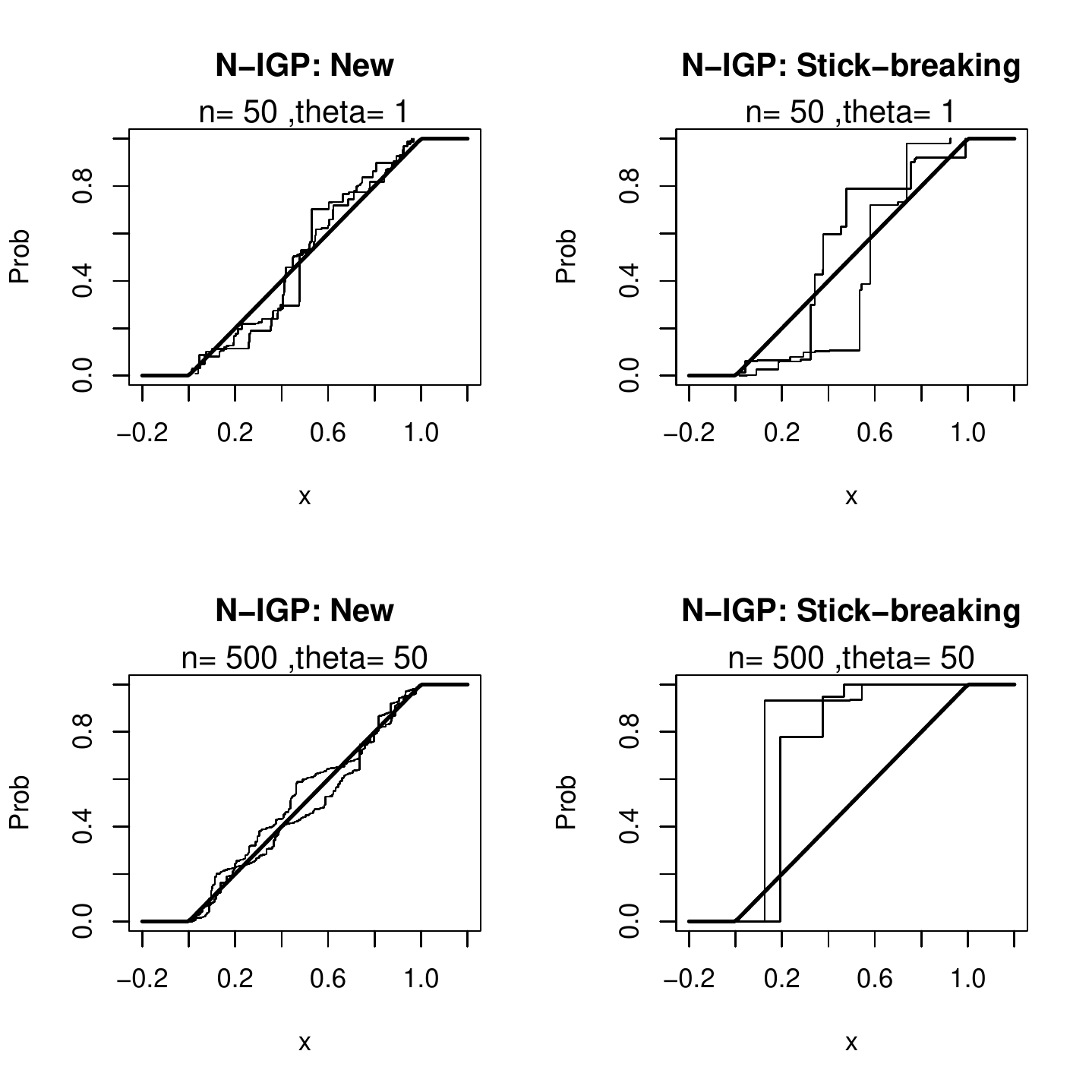}
\caption{Sample paths of the normalized inverse-Gaussian process $P_{H,\alpha,\theta}$, where
$H$ is the uniform distribution on $[0,1]$, $\theta=1,50$. The solid line  denotes the  cumulative distribution function  of $H$.}
%\vspace{30pt}
\end{figure}

\section{Acknowledgments.} The research of the authors is supported by research grants from the \textbf{Natural Sciences and
Engineering Research Council of Canada (NSERC)}.

%%%%%%%%%%%%%%%%%%%%%%%%%%%%%%%%%%%%%%%%%%%%%%%%%%%%%%%%%%%%%%%%%%%%%%%%%%%%%%%%%%%%%%%%%%%%%%%%%%%%%%%%%%%%%%%%%%%%%%%%%%%%

\begin{thebibliography}{}

\bibitem{AS} Abramowitz, M., and Stegun, I.A. (1972). \emph{Handbook of Mathematical Functions with Formulas, Graphs, and Mathematical Tables}. New York, Dover.

\bibitem{BL} Balakrishnan, N., and Lai, C.  (2009). \emph{Continuous Bivariate Distributions}. Springer-Verlag.


\bibitem{BH} Buntine, W. and Hutter, M. (2010). A Bayesian review of the Poisson-Dirichlet process. http://arxiv.org/abs/1007.0296.


\bibitem{C1} Carlton, M.A.  (1999). Applications of the two-parameter Poisson-Dirichlet distribution.
Unpublished Ph.D. thesis, Department of Statistics, University of California, Los Angeles.

\bibitem{FLM} Favaro, S., Lijoi, A., Mena, R., and Pr\"unster, I. (2009). Bayesian nonparametric
inference for species variety with a two parameter Poisson-Dirichlet process prior.
\emph{Journal of the Royal Statistical Society Series B}, {71}, 993-1008.

\bibitem{Feng} Feng, S. (2010). \emph{The Poisson-Dirichlet Distribution and Related Topics}. Springer-Verlag, New
York.

\bibitem{FK} Ferguson, T. S., and Klass, M. J. (1972). A Representation of Independent Increment Processes without Gaussian Components. \emph{The Annals of
    Mathematical Statistics}, {1}, 209-230.

\bibitem{F1}  Ferguson, T.S. (1973). A Bayesian Analysis of Some Nonparametric Problems. \emph{The Annals of Statistics}, {{1}}, 209-230.

\bibitem{IJ} Ishwaran, H., and James, L.F. (2001).  Gibbs Sampling Methods for Stick-Breaking Priors. \emph{Journal of the
American Statistical Association}, {96}, 161-173.

\bibitem{JLL} Jang, G.H., Lee, J. and Lee, S. (2010). Posterior consistency of species sampling priors. \emph{Statistica Sinica}, {20}, 581-593.

\item Kottas, A., Gelfand, A.E. (2001). Bayesian semiparametric median regression modeling.
{Journal of the American Statistical Association} 96, 1458-1468.

\bibitem{LMP} Lijoi, A., Mena, R.H., and Pr\"unster, I. (2005). Hierarchical mixture modelling with normalized
inverse Gaussian priors. \emph{Journal of the American Statistical Association},  {100}, 1278-1291.

\bibitem{MT} Muliere, P., and  Tardella, L. (1998). Approximating distributions of random functionals of Ferguson-Dirichlet prior. \emph{The Canadian Journal of
    Statistics}, {26}, 283-297.

\bibitem{NQM} Navarrete, C., Quintana, F.A. and M\"uller, P. (2008). Some issues on nonparametric Bayesian modeling using
species sampling models. S\emph{tatistical Modelling}, {8}, 3-21.

\bibitem{PY} Pitman, J.  and Yor, M.  (1997). The two-parameter Poisson-Dirichlet distribution derived
from a stable subordinator. \emph{The Annals of Probability}, {2}, 855-900.

\bibitem{SE}  Sethuraman, J. (1994). A constructive definition of Dirichlet priors. \emph{Statistica Sinica}, {4}, 639-650.

\bibitem{Teh} Teh, Y.W. (2006). A hierarchical Bayesian language model based on Pitman-Yor processes. In \emph{proceedings of the
21st International Conference on Computational Linguistics and 44th Annual Meeting of the
Association for Computational Linguistics},  985-992.


\bibitem{ZA}  Zarepour, M., and Al Labadi, L. (2012). On a Rapid Simulation of the Dirichlet Process. \emph{Statistics and Probability Letters}, 82, 916-924.


\end{thebibliography}
\end{document}